\documentclass[prd,singlecolumn,showpacs]{revtex4}

\begin{document}

\title{The gravitational equation in higher dimensions}

\vskip 1.0truecm

\author{Naresh Dadhich}\email{nkd@iucaa.ernet.in}
\affiliation{Centre for Theoretical Physics, Jamia Millia Islamia,
New Delhi 110025, India} \affiliation{Inter-University Centre for
Astronomy \& Astrophysics, Post Bag 4, Pune 411 007, India}

\vskip 1.5truecm

\begin{abstract}
Like the Lovelock Lagrangian which is a specific homogeneous polynomial in Riemann curvature, for an alternative derivation of the gravitational
equation of motion, it is possible to define a specific homogeneous polynomial analogue of the Riemann curvature, and then the trace of its
Bianchi derivative yields the corresponding polynomial analogue of the divergence free Einstein tensor defining the differential operator for
the equation of motion. We propose that the general equation of motion is $G^{(n)}_{ab} = -\Lambda g_{ab} +\kappa_n T_{ab}$ for $d=2n+1, \,
2n+2$ dimensions with the single coupling constant $\kappa_n$, and $n=1$ is the usual Einstein equation. It turns out that gravitational
behavior is essentially similar in the critical dimensions for all $n$. All static vacuum solutions asymptotically go over to the Einstein
limit, Schwarzschild-dS/AdS. The thermodynamical parameters bear the same relation to horizon radius, for example entropy always goes as
$r_h^{d-2n}$ and so for the critical dimensions it always goes as $r_h, \, r_h^2$. In terms of the area, it would go as $A^{1/n}$. The
generalized analogues of the Nariai and Bertotti-Robinson solutions arising from the product of two constant curvature spaces, also bear the
same relations between the curvatures $k_1=k_2$ and $k_1=-k_2$ respectively.
\end{abstract}


\maketitle

\section{Introduction}

What stands gravity apart from rest of the physics is its universal character that it links to everything including massless particles and hence
it can only be described by the spacetime curvature, and its dynamics has therefore to follow from the geometric properties of the Riemann
curvature tensor \cite{n1}. The Einstein gravitational equation could be deduced from the geometric property of Riemann curvature, known as the
Bianchi identity, implying vanishing of its Bianchi derivative identically. Then on taking its trace yields the divergence free second rank
symmetric Einstein tensor, which defines the differential operator on the left for the equation while on the right gravitational source, energy
momentum distribution described by a second rank symmetric tensor with the condition of vanishing divergence. This is the case for Einstein
gravity which is linear in Riemann curvature, and its vacuum is trivially flat in $3$ dimension and it becomes dynamically non-trivial in $4$
dimension. \\

The question is, could this be generalized to a polynomial analogue of the Riemann tensor? Consider a tensor with the same symmetry properties
as the Riemann which is a homogeneous polynomial of degree $n$ in Riemann, and then demand that the trace of its Bianchi derivative vanishes.
This will fix the coefficients in the polynomial and will give the divergence free second rank symmetric tensor $G^{(n)}_{ab}$, the $n$th order
analogue of the Einstein tensor, which is the same as what one would get from the variation of the $n$th order Lovelock Lagrangian
\cite{bianchi}. Thus we have the generalized polynomial Riemann curvature, $R^{(n)}_{abcd}$, which would describe gravitational dynamics in
$d=2n+1, 2n+2$ in the same manner as Riemann does for $d=3, 4$. We can define corresponding vacuum as $R^{(n)}_{ab}=0$, would it also be trivial
in $d=2n+1$ dimension? The answer is indeed, yes \cite{dgj}. It would be $R^{(n)}_{abcd}$ flat but not Riemann flat, and for that it would
describe a global monopole \cite{bv}. \\

What should be the gravitational equation in dimension $>4$? Should
it continue to be the Einstein equation which is linear in Riemann
or should it include the one following from the higher order
Riemann, $R^{(n)}_{abcd}$ yet giving the second order quasi-linear
equation? A general abiding principle is that the equation be second
order quasi-linear so that the initial value problem is well
formulated giving unique evolution. This uniquely identifies the
Lovelock polynomial Lagrangian or equivalently the above discussed
polynomial Riemann curvature \cite{bianchi}. Should all orders that
are non-trivial in the equation be included like the linear
Einstein, quadratic Gauss-Bonnet, and so on, or the only highest
one? one gi and all orders of Riemann should be included that make
non-zero contribution in the equation. Should it be $\sum
G^{(n)}_{ab}$ or $G^{(n)}_{ab}$? In the former, the each order will
have its own coupling and so there would be $n$ of them, and there
is no obvious way to fix them. Since there is only one force which
allows determination of only one coupling parameter by
experimentally measuring its strength, gravity should therefore have
only one dimensional coupling parameter and its dimension would
however depend upon the spacetime dimension. Thus we propose the
gravitational equation should in general be written as
\begin{equation}
G^{(n)}_{ab} = - \Lambda g_{ab} + \kappa_n T_{ab}
\end{equation}
for $d=2n+1, 2n+2$ dimensions. Note that $\Lambda$, which characterizes dynamics free spacetime, is part of the structure of spacetime on the
same footing as the velocity of light \cite{n2}. In what follows we wish to demonstrate that this equation imbibes beautifully the general
vacuum character \cite{dgj} as well as the static vacuum solutions asymptotically go over to the right Einstein limit, even though the linear
Einstein term is not included. This means higher order terms in curvature are only pertinent to the high energy end near the black hole horizon
while their effect weans out asymptotically at the low energy end approximating to the linear order Einstein solution, Schwarzschild-dS/AdS in
$d$ dimension \cite{n3, dpp1}. It is remarkable that the thermodynamical parameters, temperature and entropy bear universal relation to the
horizon radius for static black holes in $d=2n+1, 2n+2$, and interestingly this property also marks the characterization of this class of black
holes \cite{dpp2, dpp1}. \\

\section{The Lovelock curvature polynomial and the equation of motion}
Following Ref. \cite{bianchi}, we define the Lovelock curvature polynomial
\begin{eqnarray}\nonumber
R^{(n)}_{abcd} &=& F^{(n)}_{abcd}  - \frac{n-1}{n(d-1)(d-2)} F^{(n)} (g_{ac}g_{bd} - g_{ad}g_{bc}), \\
F^{(n)}_{abcd} &=& Q_{ab}{}{}^{mn} R_{cdmn}
\end{eqnarray}
where
\begin{eqnarray}\nonumber
Q^{ab}{}{}_{cd} &=& \delta^{a b a_1 b_1...a_n b_n}_{cdc_1 d_1...c_n d_n} R_{a_1 b_1}{}{}^{c_1 d_1},\ldots,R_{a_{n-1} b_{n-1}}{}{}^{c_{n-1}d_{n-1}}, \\
Q^{abcd}{}{}{}{}_{; d}&=&0.
\end{eqnarray}
It follows that the trace of the Bianchi derivative yields the divergence free $G^{(n)}_{ab}$;i.e.
\begin{equation}
g^{ac}g^{bd}R^{(n)}_{abcd;e} = G^{(n)}{}^{b}{}_{e;b} = 0
\end{equation}
where the analogue of $n^{th}$ order Einstein tensor is given by
\begin{equation}
G^{(n)}_{ab} = n(R^{(n)}_{ab} - \frac{1}{2} R^{(n)} g_{ab}).
\end{equation}
Note that
\begin{equation}
 R^{(n)} = \frac{d-2n}{n(d-2)}F^{(n)}
\end{equation}
which vanishes for $D=2n$ while $F^{(n)}$, the Lovelock action polynomial, is non-zero but its variation, $G^{(n)}_{ab}$ vanishes identically.
Since $R^{(n)}= g^{ab}R^{(n)}_{ab}=0$ for $d=2n$ for arbitrary $g_{ab}$, it implies $R^{(n)}_{ab}=0$ identically as it involves apart from the
metric its first and second derivatives which are arbitrary. \\

Since $G^{(n)}_{ab}$ is divergence free, we could write
\begin{equation}
G^{(n)}_{ab} = \kappa_n T_{ab} -\Lambda g_{ab}, \, \, T^{ab}_{;b}=0.
\end{equation}
This is the gravitational equation for $d=2n+1, 2n+2$ dimensions with $\kappa_n$ as the gravitational constant, and $n=1$ is the Einstein
equation for $3$ and $4$ dimensions. What degree of polynomial in Riemann should the equation have is thus determined by the spacetime
dimension. It is linear for $3, 4$, quadratic for $5, 6$, and so on. \\

\section{Universal features}

The first universal feature studied was that of gravitational field inside a uniform density sphere and it was shown that it was always given by
the Schwarzschild interior solution in Einstein as well as in Einstein-Gauss-Bonnet/Lovelock theories \cite{dmk}. Here we shall consider the
cases of static black holes, and product spaces describing the Nariai and Bertotti-Robinson spacetimes. \\

\subsection{Static black holes} The static spherically symmetric solution of the vacuum equation (1) is given by
\begin{equation}
g_{tt} = -1/g_{rr} = V = 1 - r^2(\Lambda + M/r^{d-1})^{1/n}
\end{equation}
which asymptotically takes the form of the Schwarzschild-dS/AdS solution in $d$ dimension showing the correct Einstein limit. The solution for
the general case of the Einstein-Lovelock equation can also be written in terms of the $n$th order algebraic polynomial equation which cannot be
solved in general for $n>4$. It is therefore clear that we cannot carry on with arbitrarily large number of coupling parameters. For the case of
dimensionally continued black holes \cite{btz}, it was proposed that all the couplings are determined in terms the unique ground state
$\Lambda$, and the solution is then given by $V= 1 - r^2\Lambda - M/r^{d-1/2}$ which clearly does not go over to the Einstein solution for large
$r$. This corresponded to the algebraic polynomial being degenerate. It turns out that the proper Einstein limit could be brought is simply by
considering the polynomial to be derivative degenerate \cite{dpp1}. Then the solution agrees near the horizon with the dimensionally continued
black hole and asymptotically to the proper Einstein limit, and it is the solution of the equation (1). \\

Further the thermodynamical parameters, temperature and entropy bear the universal relation to the horizon radius for the critical $d=2n+1, \,
2n+2$ dimensions \cite{dpp2}. For instance, the entropy always goes as $r_h^{d-2n}$ which for the critical dimensions would always go as $r_h,
\, r_h^2$. In terms of the area, it would however go as $A^1/n$, and hence the entropy is proportional to area only for the $n=1$ Einstein
theory. Interestingly this universality is also the characterizing property of this class of pure Lovelock black holes \cite{dpp1, dpp2}.

We would like to conjecture that the above universality property would also be true for the rotating black hole solution as and when it is
found. \\

\subsection{Product spaces: Nariai and Bertotti-Robinson solutions}

The Nariai and Bertotti-Robinson solutions arise as product of two constant curvature spaces. When the two curvatures are equal, $k_1=k_2$, it
is the Nariai solution of the equation (1) with $T_{ab}=0$ for $n=1$, and when the curvatures are equal and opposite, $k_1=-k_2$, it is the
Bertotti-Robinson solution describing the uniform electric field. The former is the $\Lambda$ vacuum spacetime but is not conformally flat while
the latter is the Einstein-Maxwell solution for uniform electric field which is conformally flat. It turns out the generalized pure Lovelock
solutions of the equation (1) for any $n$ bear out the same curvature relations for the Nariai vacuum ($k_1=k_2$) and Bertotti-Robinson uniform
electric field ($k_1=-k_2$), and the condition for conformal flatness is also $k_+k_2=0$ \cite{dp}. \\

In $d=2n+2$ dimension, we have the following general relation connecting the two curvatures, $\Lambda$ and the electric field $E$,
\begin{equation}
(k_1+k_2)E^2 = -4(k_1-k_2)\Lambda.
\end{equation}
This clearly indicates $k_1=k_2$ for $E=0$, the Nariai vacuum spacetime and $k_1=-k_2$ for $\Lambda=0$, the Bertotti-Robinson uniform electric
field spacetime. The metric is given by
\begin{equation}
ds^2 = (1-k_1r^2) dt^2 - \frac{dr^2}{1-k_1r^2} - \frac{1}{k_2}d\Sigma^2_{(d-2)}.
\end{equation}

\section{Discussion}

We have proposed that the equation (1) is the proper equation for gravity in higher dimensions. The correct equation should have the following
properties: (a) it should be second order quasi-linear, (b) for a given dimension, it should be of degree $n=[(d-1)/2]$ in the Riemann
curvature, (c) it should have only one coupling constant which could be determined by experimentally measuring the strength of the force and (d)
since higher order curvature contributions are the high energy corrections to the linear order in Riemann Einstein gravity which should wean out
asymptotically, hence solutions should tend to the corresponding Einstein solution for large $r$. The proposed equation satisfies all these
properties. The latter feature of the asymptotic Einstein limit is verified for the static black hole solutions which is however also true for
the Einstein-Gauss-Bonnet black hole. What is remarkable here is that the equation is free of the Einstein term, yet asymptotically solutions go
over to the proper Einstein limit. This means high energy effects which come through the higher order curvature terms are fully and properly
taken care by the highest order $n=[(d-1)/2]$ term, and they could be realized only in higher dimensions \cite{high-d}. It is interesting that
gravity asks for higher dimensions for realization of its high energy effects. This is because inclusion of higher orders in Riemann curvature
and the demand that the equation continues to be second order quasi-linear naturally lead to higher dimensions. This does not happen for any
other force that one has to consider higher dimension for realization of its high energy corrections. It happens for gravity because the
spacetime curvature is the basic field variable, and hence high energy effects involve higher orders in it and their contribution in the
equation, if it continues to retain its second order quasi-linear character, can be realized only in higher dimensions \cite{high-d}. We would
like to emphasize that higher dimensions and high energy effects seem to be intimately connected. Since high energy effects ask for higher
dimensions, quantum gravity should also involve higher dimensions. This is because quantum gravity should approach the classical limit via
the high energy intermediate limit. \\

One of the problems with the Einstein-Lovelock solutions is the number of coupling constants and there is no way to fix them. For the
dimensionally continued static black holes, all the couplings were prescribed in terms of the unique ground state $\Lambda$ \cite{btz}. These
solutions were however not asymptotically Einstein, Schwarzschild-dS/AdS. Instead the corresponding solutions of the equation (1) have the right
limits at both ends, nearer to horizon agreeing with the dimensionally continued and asymptotically to Schwarzschild-dS/AdS.  This is indicative
of the inherent correctness of the equation. The universal character of gravity in the critical dimensions is another very attractive feature of
the equation. That the vacuum, $G^{(n)}_{ab} = 0$, in the odd critical dimension is always trivial, $R^{(n)}_{abcd} = 0$ \cite{dgj}. All this
taken together points to the fact that the equation (1) is the right equation for gravitation in higher dimensions. \\

For a given order $n$ in the Riemann curvature, the critical dimensions are $d=2n+1, \, 2n+2$ and it is trivial/kinematic in the former and it
becomes dynamic in the latter. This is a universal general feature. In the critical dimensions, gravity has the similar behavior as indicated by
universality of the thermodynamic parameters in terms of the horizon radius and of the Nariai and Bertotti-Robinson solutions. It is interesting
to note that in terms of black hole area, entropy is always proportional to $A^{1/n}$ and so it is proportional to area only for the $n=1$
Einstein gravity. This is an interesting general result that entropy always goes as the $n$th root of area of the black hole. In an intuitive
sense we can say that it is $n$th root of the Einstein gravity for the critical $d=2n+1, \, 2n+2$ dimensions.  \\

All this we have established for the simple case of static black hole but we believe that it is indeed a general feature and hence should be
true for the stationary rotating black hole as well. So far there exists no rotating pure Lovelock black hole solution, and this conjecture
would be verified as and when a solution is found. \\

\section*{Acknowledgement} It is a pleasure to thank the organizers of the conference, Relativity and Gravitation: 100 years after Einstein in
Prague, June 25-28, 2012.

\end{document}